\documentclass[aps,
prl,
twocolumn,
reprint,
noeprint,
superscriptaddress,
amsmath,
amssymb,
]{revtex4-2}
\usepackage[english]{babel}
\usepackage{amsmath}
\usepackage{graphicx}
\usepackage[colorlinks=true, allcolors=blue]{hyperref}
\usepackage{times}
\usepackage{ulem}
\usepackage{pdfpages}

\makeatletter
\AtBeginDocument{\let\LS@rot\@undefined}
\makeatother

\begin{document}

\title{Enhancing Dynamic Range of Sub-Standard-Quantum-Limit Measurements via Quantum Deamplification}

\author{Qi Liu}
\affiliation{Department of Physics, MIT-Harvard Center for Ultracold Atoms and Research Laboratory of Electronics, Massachusetts Institute of Technology, Cambridge, Massachusetts 02139, USA}

\author{Ming Xue}
\thanks{Current address: Department of Physics, Nanjing University of Aeronautics and Astronautics, Nanjing 211106, China.}
\affiliation{State Key Laboratory of Low Dimensional Quantum Physics, Department of Physics, Tsinghua University, Beijing 100084, China}

\author{Matthew Radzihovsky}
\affiliation{Department of Physics, MIT-Harvard Center for Ultracold Atoms and Research Laboratory of Electronics, Massachusetts Institute of Technology, Cambridge, Massachusetts 02139, USA}

\author{Xinwei Li}
\thanks{Current address: Beijing Academy of Quantum Information Sciences, Xibeiwang East Road, Beijing 100193, China}
\affiliation{State Key Laboratory of Low Dimensional Quantum Physics, Department of Physics, Tsinghua University, Beijing 100084, China}

\author{Denis V. Vasilyev}
\affiliation{Institute for Theoretical Physics, University of Innsbruck, 6020 Innsbruck, Austria}

\author{Ling-Na Wu}
\email[]{lingna.wu@hainanu.edu.cn}
\affiliation{Center for Theoretical Physics \& School of Physics and Optoelectronic Engineering, Hainan University, Haikou 570228, China}

\author{Vladan Vuleti\'{c}}
\affiliation{Department of Physics, MIT-Harvard Center for Ultracold Atoms and Research Laboratory of Electronics, Massachusetts Institute of Technology, Cambridge, Massachusetts 02139, USA}

\date{\today}
\begin{abstract}
Balancing high sensitivity with a broad dynamic range is a fundamental challenge in measurement science, as improving one often compromises the other. 
While traditional quantum metrology has prioritized enhancing local sensitivity, a large dynamic range is crucial for applications such as atomic clocks, where extended phase interrogation times contribute to wider phase range. 
In this Letter, we introduce a novel quantum deamplification mechanism that extends dynamic range at a minimal cost of sensitivity. Our approach uses two sequential spin-squeezing operations to generate and detect an entangled probe state, respectively. 
We demonstrate that the optimal quantum interferometer limit can be approached through two-axis counter-twisting dynamics. 
Further expansion of dynamic range is possible by using sequential quantum deamplification interspersed with phase encoding processes.
Additionally, we show that robustness against detection noise can be enhanced by a hybrid sensing scheme that combines quantum deamplification with quantum amplification. Our protocol is within the reach of state-of-the-art atomic-molecular-optical platforms, 
offering a scalable, noise-resilient pathway for entanglement-enhanced metrology.
\end{abstract}
\maketitle

{\it Introduction.}---
In the field of precision measurement, achieving high sensitivity to the signal of interest across a broad measurement range is a key goal~\cite{degen17rmp,Luca18rmp,huang24entanglement}. 
Considerable advancements have been made in surpassing the standard quantum limit~(SQL)---the best phase sensitivity achievable with uncorrelated particles---by leveraging entangled probe states,
including squeezed states~\cite{Ueda1993PRA,Kuzmich97Spin,Hald99spin,Gross10nonlinear,Riedel2010atom,Smerzi2013PRL,Roman16Bell,Polzik16Entanglement,Bohnet16quantum,colombo2022entanglement,mao2023quantum,greve2022entanglement,LIGO23PRX,LIGO24Science, Robinson24direct},
Dicke states~\cite{Huegla1997,Lucke11twin,Zhang_2014,Luo17deterministic,zou2018beating}, 
and Schr\"odinger-cat-like states~\cite{Sylvain18NC,cao2024multi,finkelstein2024universal,kielinski2024ghz,Yang2024minutes,liu2024}. 
However, the enhanced sensitivity obtained in this way is usually confined to a narrow phase range~\cite{zollerPRX2021,Kaubruegger23Optimal,vasilyev2024optimal}. 
Moreover, the detection of entangled states is susceptible to technical noise, which  diminishes the theoretical metrological benefits promised by entanglement~\cite{Lukin04stability,schulte2020prospects}.
To address the latter challenge, quantum amplification (QA) techniques~\cite{davis16approaching,Frowis16Detecting, nolan17optimal,Haine2018PRA,liu23cyclic,hu23spin,hu2023nonlinear} have been developed.
These methods typically amplify the signal through squeezing-unsqueezing protocols~\cite{hosten2016quantum,Linnemann16Quantum,Burd19Science,colombo2022time,liu2022nonlinear,mao2023quantum,cao23detection}. 
The introduction of unsqueezing,
however, comes with the trade-off of a limited phase range within which beyond-SQL sensitivity can be achieved~\cite{Simone22APLreview}.

\begin{figure}
    \centering
    \includegraphics[width=1\linewidth]{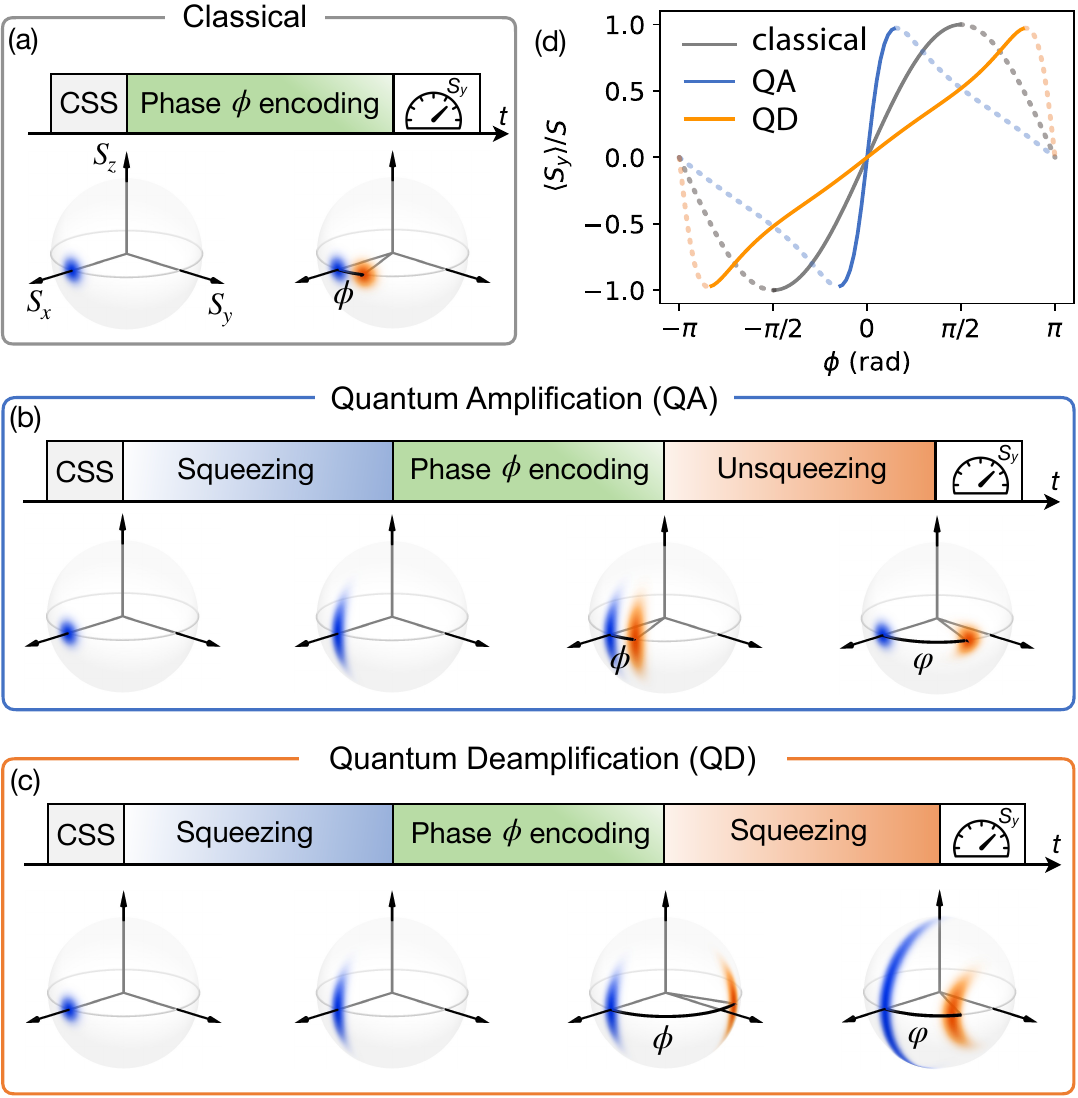}
    \caption{The interferometry time sequence and corresponding Husimi representations of states for (a) classical probe, (b) quantum amplification (QA) , and (c) quantum deamplification (QD).
    The blue (orange) shading denotes the state without (with) phase $\phi$ encoding.  
    In the QA and QD protocols, $\varphi$ denotes the phase after amplification or deamplification, respectively.
    (d) Normalized spin observable $\langle S_y\rangle/S$ as a function of phase $\phi$. The coherent spin state provides a dynamic range around $[-\pi/2,\pi/2]$ (grey solid line), within which the phase can be distinguished unambiguously.
    QA narrows the dynamic range (blue solid line), while QD extends the dynamic range to nearly $[-\pi,\pi]$ (orange solid line). Dashed lines denote the intervals where phase slip error occurs. Calculations are performed in the TACT model with squeezing or unsqueezing time of $t=0.015$. $N = 100$ and $\chi = 1$ are used in this paper unless otherwise indicated.}
    \label{fig1}
\end{figure}

While the emphasis has predominantly been on boosting local sensitivity, the demand for a broad measurement range in practical applications is equally important and warrants greater attention. {In the scenario of atomic clocks~\cite{RevModPhys.87.637}, a broader measurement range enables longer phase interrogation times, which can reduce frequency instability even if local sensitivity is slightly compromised.} Previous studies have shown that by using auxiliary interferometers, one can enhance the {\it dynamic range}, within which the phase can be extracted from the measurement unambiguously. For instance, employing dual-quadrature measurement with two ensembles doubles dynamic range~\cite{rosenband2013exponentialscalingclockstability,Li22Improved}. Even greater enhancement in dynamic range is possible through the use of multiple ensembles with different interrogation phases~\cite{Hume16probing,Borregaard13near,Dimitry20Atom,shaw23multiensemble,zheng24reducing}. By integrating these techniques with entangled states, it is possible to achieve beyond-SQL phase sensitivity across an extended phase range~\cite{Li22Improved,cao2024multi,finkelstein2024universal,direkci2024heisenberg,liu23fullperiod,zhou2024high}. The ultimate goal is to optimize both estimation precision and dynamic range by leveraging quantum resources native to the experimental platform. This can be achieved via quantum variational optimization, which designs quantum circuits generating optimal input states and measurements that minimize the Bayesian mean squared error for a given prior distribution of a parameter~\cite{zollerPRX2021,ThurtellPRR2024}. Experimental implementations~\cite{marciniak2021optimal} of such circuits approach the performance of an optimal quantum interferometer (OQI)~\cite{Macieszczak2014} but face increasing circuit complexity as system size grows. The challenge lies in identifying quantum sensing protocols and resources that enable simple, robust circuits capable of nearing OQI performance despite experimental imperfections.

In this paper, we introduce a novel mechanism that leverages {\it quantum deamplification}~(QD) to extend dynamic range using a single ensemble in the presence of detection noise. 
Unlike the QA protocols~\cite{davis16approaching,hosten2016quantum,colombo2022time}, 
our method employs two sequential squeezing stages to deamplify the encoded phase at a minimal cost of local sensitivity, 
thereby achieving a broader dynamic range. 
We demonstrate this mechanism by using a generalized Ramsey interferometer based on two-axis counter-twisting (TACT) interaction~\cite{Ueda1993PRA,luo2024hamiltonianengineeringcollectivexyz,miller24tact}, 
which is both efficient in generating spin squeezing and particularly elegant for illustrating the core idea of our approach.
This method approaches the OQI limit
in sensing performance using only one TACT gate each for state preparation and measurement. Furthermore, interspersing phase encoding with TACT squeezing---progressively deamplifying the signal---expands dynamic range further, albeit with reduced local sensitivity, bringing the overall single-shot sensing performance closer to the OQI limit. To enhance noise robustness, we propose a hybrid sensing protocol combining QD, QA, and adaptive one-way local operations and classical communication measurements~\cite{Chitambar2014}, enabling high-sensitivity phase estimation over a wide dynamic range despite experimental imperfections.

{\it Quantum (de)amplification.}---
We consider estimation of the phase $\phi$
in an atomic interferometer consisting of $N$ identical two-level atoms.
For a classical probe in the coherent spin state (CSS) with an isotropic quasiprobabilty distribution, 
as shown in Fig.~\ref{fig1}(a),
the measurement precision is inherently bounded by the SQL, \( 1/\sqrt{N} \)~\cite{ma2011quantum,Luca18rmp}.
The ambiguity-free dynamic range is confined to \([-\pi/2, \pi/2] \),
within which the observable varies monotonically with $\phi$ [grey solid curve in Fig.~\ref{fig1}(d)]. This limitation arises because \( \phi \) and \( \pm\pi - \phi \) 
yield the same measurement outcome, leading to significant bias error when the phase is out of the dynamic range.
We denote the phase after (de)amplification in QA (QD) as $\varphi$.
In the QA protocol [Fig.~\ref{fig1}(b)], 
a squeezing-encoding-unsqueezing sequence amplifies the encoded signal. 
Since the amplified phase $\varphi$ can only be distinctly resolved within the range $[-{\pi}/{2}, {\pi}/{2}]$, the estimation range of encoded phase $\phi$~($\le\!\varphi$) is further limited.
This is illlustrated by the blue solid curve in Fig.~\ref{fig1}(d),
which shows a steeper gradient near small phases and a reduced dynamic range.
In contrast, we propose introducing a squeezing instead of unsqueezing process before measurement, 
which deamplifies the signal $\phi$ [see Fig.~\ref{fig1}(c) for the QD sequence].
Given that the sensing range of the deamplified phase 
$\varphi$ is confined to the classical boundary of
$[-{\pi}/{2}, {\pi}/{2}]$, 
the ambiguity-free dynamic range of the encoded phase $\phi$~($\ge\!\!\varphi$) is extended beyond this interval,
as shown by the orange solid curve in Fig.~\ref{fig1}(d).

\begin{figure}
    \centering
     \includegraphics[width=1\linewidth]{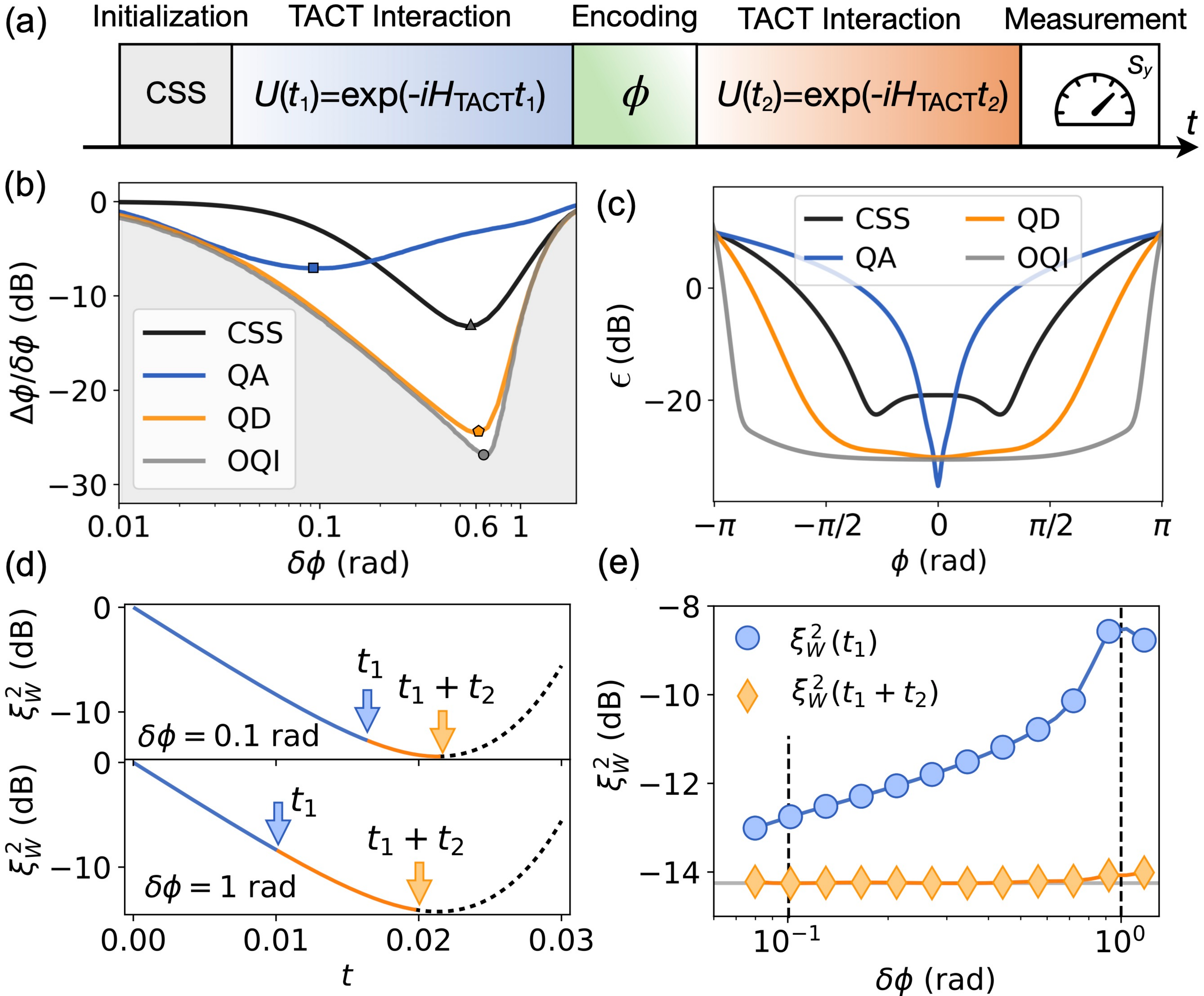}
    \caption{ (a) QD-based sensing with TACT interactions.
    The protocol sequence starts from a coherent spin state (CSS) along $x$ direction which evolves to a spin squeezed state under TACT evolution $U(t_1)$.
    After a phase ($\phi$) encoding operation, $U(t_2)$ deamplifies the signal before the final measurement of $\hat S_y$. 
    (b) The sensor performances, characterized by $\Delta\phi/\delta\phi$ [plotted in dB as $20\log_{10}(\Delta\phi/\delta\phi)$], 
    of OQI~(grey line) and QD sensors (orange line) are optimized at each $\delta\phi$,  while QA-based sensor (blue line) with TACT interaction uses a fixed interacting time ($t_1=-t_2=0.022$) corresponding to the optimal spin squeezing.
    The grey shading denotes the inaccessible regime with one-time phase encoding, bounded by the OQI limit. 
    Markers highlight the best performance (minima) at the turning point.
    (c) MSE of the sensing schemes, evaluated at $\delta\phi$ and TACT squeezing times corresponding to the minima (markers) of $\Delta\phi/\delta\phi$ in (b). 
    (d) TACT-interacting times $t_1$ (blue) and $t_1+t_2$ (orange) and their corresponding Wineland parameters ($\xi_W^{2}$) evolution, 
    for two selected $\delta\phi$ marked by dashed vertical lines in (e).
    (e) Wineland parameters for the $\delta\phi$-dependent optimal probe (blue circles) and final state (orange diamonds) in the QD scheme. The horizontal grey line marks the optimal spin squeezing obtainable with TACT interaction.}
    \label{fig2}
\end{figure}

{\it QD-based sensing with TACT interactions.}---
Building on the above discussion, we now delve into a more detailed examination of our proposal. 
We employ the TACT squeezing model~\cite{Ueda1993PRA}, governed by the Hamiltonian $\hat{H}_{\rm TACT}=-\chi(\hat{S}_y\hat{S}_z+\hat{S}_z\hat{S}_y)$ with interaction strength $\chi$, where $\hat{S}_{x,y,z}=\sum_{k=1}^N \hat\sigma_{x,y,z}^{(k)}/2$ are the collective spin operators
and $\hat\sigma_{x,y,z}^{(k)}$ the Pauli operators for the $k$-th atom.
The structure of our QD-based quantum interferometer is shown in Fig.~\ref{fig2}(a). 
It starts from a CSS $|\Psi_0\rangle$, with all spins polarized along the $x$-direction,
followed by the application of the TACT interaction twice [also cf. Fig.~\ref{fig1}(c)]. 
The first instance of TACT interaction precedes the phase encoding process, serving to prepare an entangled probe state, 
while the second application takes place after phase encoding, facilitating interaction-based readout~\cite{nolan17optimal,Haine2018PRA,liu23cyclic,cao23detection}.
The encoded phase is inferred from the collective spin $\hat S_y$ through projective measurements.

The phase estimation accuracy is characterized by the mean squared error (MSE) with respect to the actual phase $\phi$, $\epsilon(\phi)= \sum\nolimits_m{[\phi-\phi_{\rm est}(m)]^2 p(m|\phi)},$
where $p(m|\phi)$ denotes the conditional probability of measurement outcome $m$,
and is given by
$p(m|\phi) = |\langle m|\hat{U}(t_2)e^{-i\phi \hat{S}_z}\hat{U}(t_1)|\Psi_0\rangle|^2$,
with $|m\rangle\equiv|S=N/2,S_y=m\rangle$ being the eigenstate of $\hat S_y$ with eigenvalue $m$. 
Here, $\hat{U}(t)=e^{-i\hat{H}_{\rm TACT}t}$, 
and $t_1$, $t_2$ represent the evolution time parameters for the two TACT segments.  
The optimal values of $t_1$ and $t_2$ are to be found through optimization. 
We assume a linear estimator $\phi_{\rm est}(m)=am$, 
and the optimal $a$ is determined from the measurement outcome distribution~\cite{ThurtellPRR2024}.
Our goal is to achieve optimal sensitivity for phase $\phi$ in a prior phase range $\delta\phi$. 
Assuming a Gaussian prior phase distribution centered at zero, 
${\cal P}_{\delta \phi}(\phi) = \exp{[-{\phi^2}/{(\sqrt{2}\delta\phi)^2}]/{\sqrt{2\pi(\delta\phi)^2}}},$
we minimize the MSE averaged over the prior distribution,
defining the Bayesian mean squared error (BMSE):
$(\Delta \phi)^2 = \int{d\phi \;\epsilon(\phi){\cal P}_{\delta \phi}(\phi)}.$

Figure~\ref{fig2}(b) and (c) showcase the performance of our QD-based sensor, 
compared to the OQI, classical, and QA-based sensor. 
In Fig.~\ref{fig2}(b), we depict the ratio of posterior uncertainty $\Delta\phi$ to the prior uncertainty $\delta \phi$, 
minimized individually at each $\delta \phi$ for QD sensor and OQI.
This quantity is related to frequency  stability in  atomic clock, characterized by the Allan deviation~\cite{Luca20Heisenberg,marciniak2021optimal}. 
A smaller value of $\Delta\phi/\delta\phi$ indicates a lower phase estimation error (smaller $\Delta\phi$) over a broader phase range (larger $\delta\phi$).
For all cases, $\Delta\phi/\delta\phi$ initially declines as $\delta\phi$ grows 
and then starts to rise due to increased probability of phase slip error~\cite{zollerPRX2021,ThurtellPRR2024,schulte2020prospects}.
When compared to the classical Ramsey interferometer~(black line, CSS), 
the QA-based sensor~(blue line, QA) enhances sensitivity only at  small $\delta\phi$ values, 
due to reduced dynamic range caused by signal amplification.
The QD-based sensor~(orange curve), involving a squeezing-encoding-squeezing sequence, significantly enhances the precision of phase estimation at large $\delta\phi$.
Notably, its performance nearly approaches the theoretical limit given by the OQI (grey curve) with the optimal probe state and measurement~\cite{Macieszczak2014}.
Figure~\ref{fig2}(c) displays the MSE as a function of the phase $\phi$. 
The prior uncertainties $\delta\phi$ and TACT squeezing times are selected to achieve the minimal $\Delta\phi/\delta\phi$ as indicated by the markers in panel (b). 
It is evident that the QA-based sensor~(blue curve) outperforms the classical interferometer~(black curve) only for small values of $\phi$. 
In contrast, the QD-based sensing protocol (orange curve) diminishes the error throughout the entire range of $[-\pi, \pi]$.

\begin{figure}[!b]
    \centering
    \includegraphics[width=1\linewidth]{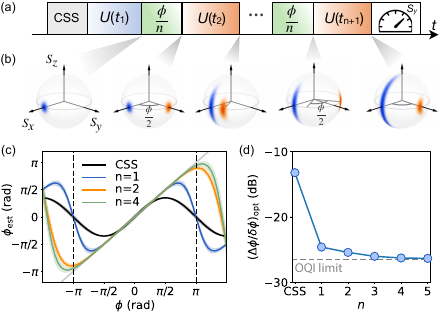}
    \caption{(a) Schematics of the sequential QD interspersed with phase encoding.
     (b) Evolution of quantum state on the Bloch spheres for $n=2$. 
     Squeezing times are $t_1=0.0022$, $t_2=0.014$, and $t_3=0.0061$, 
     with a total time cost close to the optimal squeezing time in Fig.~\ref{fig2}(d).  
    (c) Estimated phase $\phi_{\rm est}$ as a function of the real phase $\phi$. 
     The shading represents the square root of the MSE.
     The grey solid line denotes the unbiased estimation with $\phi_{\rm est}=\phi$. 
     Black dashed vertical lines label $\phi=\pm\pi$.
    (d) Optimal $\Delta\phi/\delta\phi$ as a function of $n$. The OQI limit is illustrated by the grey dashed line.
    }
    \label{fig3}
\end{figure}

The superior performance of the QD-based sensing protocol can be credited to the aforementioned squeezing-squeezing mechanism which allows for an extended dynamic range. 
This is clearly demonstrated in Fig.~\ref{fig2}(d), 
which shows the evolution of the system's squeezing degree for two distinct $\delta\phi$.
The squeezing degree is quantified by the Wineland parameter~\cite{Wineland92spin,Wineland94squeezed}, 
$\xi^{2}_W=N\Delta^2 \hat S_{\perp,\rm{min}}/|\langle \mathbf{\hat S}\rangle|^2$, 
where $\Delta^2 \hat S_{\perp,\rm{min}}$ denotes the minimal spin fluctuation on the plane perpendicular to the mean spin direction and the collective spin vector $\mathbf{\hat S}=(\hat S_x, \hat S_y, \hat S_z)$.
Figure~\ref{fig2}(d) reveals a sequential squeezing process, 
where the initial squeezing ({blue segment}) creates a spin squeezed state as a probe, 
while the subsequent squeezing~({orange segment}), leads to signal deamplification, 
thereby enhancing the dynamic range.
An overview of the Wineland parameter as a function of $\delta\phi$ is shown in Fig.~\ref{fig2}(e).
One can see that the squeezing of the probe state (blue) is always less pronounced than that of the final state (orange).  
Particularly, 
the final state is close to the optimal squeezed state. 
In other words, the journey to the optimal squeezed state is divided into two segments, 
one for state preparation and the other for signal deamplification. 
The timing for phase encoding, nestled between these two processes, 
depends on the prior phase uncertainty: for a broader prior distribution (larger $\delta\phi$), 
phase encoding should occur earlier to maximize the signal deamplification effect,
as shown in Fig.~\ref{fig2}(d).
Furthermore, we claim that the QD mechanism based on sequential squeezing is not limited to the TACT model and can be generalized to other interactions~\cite{supplementary}.


\begin{figure}[!t]
    \centering
    \includegraphics[width=1\linewidth]{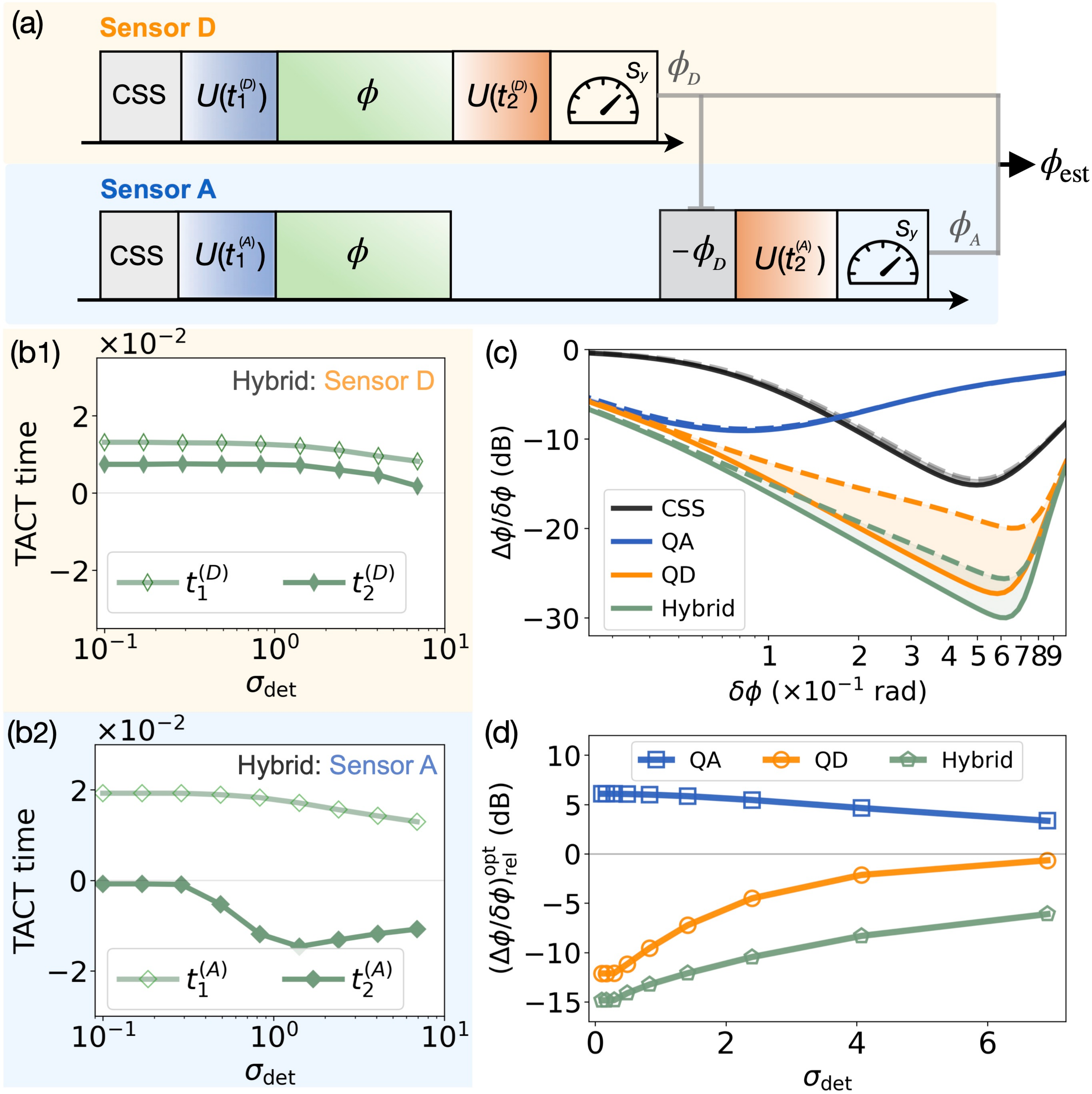}
    \caption{Hybrid sensing with adaptive measurement. 
    (a) Two sensors are conjointly used for measuring the phase of LO. 
    The phase estimated from the sensor D is subtracted from the phase encoded on sensor A. 
    (b1-b2) The optimal TACT-interacting times for the two sensors as a function of detection noise. Negative time in (b2) corresponds to a sign-flip of $\chi$.
    (c) Optimized $\Delta\phi/\delta\phi$ for various sensing schemes.
    The solid (dashed) curves denote the results in the absence (presence, with $\sigma_{\rm det}=2$) of detection noise.
    (d) The minimal phase estimation error with respect to the classical sensor,
    $(\Delta\phi/\delta\phi)^{\rm opt}_{\rm rel}\equiv(\Delta\phi/\delta\phi)^{\rm opt}/(\Delta\phi/\delta\phi)^{\rm opt}_{\rm CSS}$,
    as a function of detection noise $\sigma_{\rm det}$. }
    \label{fig:3}
\end{figure}

{\it Sequential QD interspersed with phase encoding.}
---Given the cyclical nature of phase encoding via unitary~$e^{-i\phi \hat{S}_z}$---where $\phi$ and $\phi+2\pi$ 
are indistinguishable—the maximum dynamic range of the aforementioned methodologies is inherently restricted to the interval $[-\pi,\pi]$. 
To surmount this limitation, we propose enhancing phase encoding with sequential~QD~\cite{supplementary}.
As illustrated in Fig.~\ref{fig3}(a), 
this approach involves the iterative application of 
spin squeezing interspersed with phase encoding.
In the limit of infinite iterations, it approaches concurrent entanglement generation and phase interrogation, as previously used to unify state preparation and phase encoding~\cite{simon20machine,huang23Quantum}.
This methodical layering of phase encoding and TACT interaction serves to deamplify the signal progressively~[Fig.~\ref{fig3}(b)], 
thereby extending the dynamic range beyond $[-\pi, \pi]$.
In Fig.~\ref{fig3}(c), the estimated phase $\phi_{\rm est}$ is depicted as a function of the encoded phase $\phi$,
across various iterations $n$. 
The ideal estimation is unbiased, 
aligning $\phi_{\rm est}$ with $\phi$, as indicated by the grey solid line. 
One can see that with each additional layer of phase encoding,
the unambiguous phase estimation range is enhanced.
The sequential QD also brings the benefit of reducing BMSE as shown in Fig.~\ref{fig3}(d), 
which converges to the OQI limit with increasing $n$.

{\it Hybrid sensing with adpative measurement.}---
In the absence of detection noise, the QD sensor is capable of approaching the OQI performance, 
providing a near-optimal balance between the local sensitivity and phase sensing range. 
However, the QD sequence ends up with a highly squeezed state before measurement, 
which renders the metrological performance susceptible to detection noise. 
This challenge also exists in the circuit optimization for a generalized Ramsey interferometers based on one-axis twising~\cite{zollerPRX2021,ThurtellPRR2024}, 
and poses one practical limitation of the latter in large systems.
In contrast, QA sensor exhibits strong resilience to detection noise, 
though this robustness comes at the expense of a narrower dynamic range.  
Here we study a {\it hybrid sensing} scheme~\cite{Luca20Heisenberg} to harness the complementary strengths of these two sensors in a synergistic manner. 
The QD sensor, with its ability to offer a broad dynamic range is utilized for the preliminary phase estimation. 
It allows for the sensing process to effectively handle a variety of phase shifts, 
although the sensitivity would be compromised by detection noise. 
To address this, we then employ the QA sensor with an {\it adaptive} measurement to refine the phase estimation, 
improving the system's resilience to noise and reducing the phase estimation error. 
Note that due to the extremely narrow dynamic range of the QA sensor, a reliable initial estimate over a broad range is essential~\cite{supplementary}. The QD sensor fulfills this role by providing both a wide dynamic range and sufficient precision in the preliminary phase estimation.

To validate our idea numerically, we design a hybrid system comprising two TACT-based sensors interrogated with the same local oscillator (LO), as shown in Fig.~\ref{fig:3}(a). 
The first sensor (D) provides an initial estimate of the LO phase $\phi$, called $\phi_D$.
This phase estimate is fed back to the second sensor (A) by adjusting the control pulse phase, effectively adding $-\phi_D$ to the interrogation phase $\phi$.
Finally, by analyzing its output to get $\phi_A$, 
we estimate the phase $\phi$ as $\phi_{\rm est}=\phi_D + \phi_A$.

We optimize the TACT-interacting times at two stages—before and after the phase encoding—in both sensors, 
\{$t_1^{(D)}$, $t_2^{(D)}$, $t_1^{(A)}$, $t_2^{(A)}$\}, 
with the goal of minimizing the BMSE in the presence of detection noise $\sigma_{\rm det}$. The influence of detection noise is modeled as a Gaussian convolution that transforms the initial positive-operator-valued
measurement  $|m\rangle\langle m|$ into $
|\tilde{m}\rangle\langle\tilde{m}|=\sum_{m^{\prime}} \Gamma_{m, m^{\prime}}\left|m^{\prime}\right\rangle\left\langle m^{\prime}\right|$, where $
\Gamma_{m, m^{\prime}}=e^{-\left(m-m^{\prime}\right)^2 / 2 \sigma_{\rm det}^2} / \sum_m e^{-\left(m-m^{\prime}\right)^2 / 2 \sigma_{\rm det}^2}$~\cite{Haine18Using}.
The results, depicted in Fig.~\ref{fig:3}(b), 
illustrate how the interaction times vary with the level of detection noise.
The interaction times for the first sensor are both positive ($t^{(D)}_{1,2}>0$), 
signifying its role as a QD sensor with a sequential squeezing process. 
On the other hand, the second sensor exhibits a positive interaction time prior to phase encoding ($t_1^{(A)}>0$) 
and a negative interaction time post phase encoding ($t_2^{(A)}<0$), 
which is indicative of a QA sensor with a squeezing-unsqueezing mechanism.

In Fig.~\ref{fig:3}(c), we compare the BMSE of the hybrid sensor with several individual sensing schemes. 
To guarantee a fair comparison, 
we employ the same total number of particles in all the schemes. 
The solid (dashed) curves denote the results in the absence (presence, with $\sigma_{\rm det}=2$) of detection noise.
Among the individual sensing schemes
(either split into two independent sensors of $N$ atoms each or measured independently twice with $N$ atoms~\cite{supplementary}), 
the QA sensor (blue curves) exhibits strong resilience to detection noise (highlighted by the shading area), while it falls short in BMSE. 
In contrast, the QD sensor~(orange curves) provides a small phase estimation error over a broad range of $\delta\phi$, 
but is highly susceptible to detection noise. 
The hybrid sensor~(green curves), which combines the strengths of both QD and QA sensors, 
demonstrates robustness against detection noise and maintains
low phase estimation error across a broad range of phase. 
Figure~\ref{fig:3}(d) further highlights the resilience difference 
between the sensors.
One can see that the hybrid sensor~(green curve) consistently outperforms 
the other sensors across all levels of detection noise,
and its performance degrades slower with increasing detection noise compared to the individual QD sensor.

{\it Conclusion.}---
We introduce a novel QD mechanism to extend the dynamic range of phase measurements through squeezing-encoding-squeezing operations, which is demonstrated by using a TACT-based interferometer.
The phase encoded on the squeezed probe state gets deamplified during the second squeezing stage at a minimal cost of local sensitivity,
thereby achieving a broader dynamic range and approaching the OQI limit.
This scheme is further enhanced by sequential QD interspersed with multiple phase encoding processes.
Since the QD sequence ends in a highly squeezed state before measurement, 
its metrological performance becomes susceptible to detection noise. 
To address this limitation, we propose a hybrid sensing scheme that combines the advantages of both QD and QA sensors to achieve a precise and robust phase sensing.

Our protocols are accessible with current cold atom and molecule experiments. Recent experiments have successfully demonstrated the mean-field spin dynamics of the TACT model in cavity QED systems~\cite{luo2024hamiltonianengineeringcollectivexyz} 
and ultracold polar molecules~\cite{miller24tact}, 
making the observation of spin squeezing highly promising in the near future. 
To further evaluate the feasibility of our protocol, we analyze the effects of collective and single-particle dephasing in the supplementary material~\cite{supplementary}. Our results demonstrate that the extended dynamic range of sub-standard-quantum-limit measurements remains achievable with near-future experimental setups~\cite{edwin20entanglement, ZeyangLi23science, martin23revisiting}, and the resilience to dephasing can be further enhanced through Hamiltonian engineering.
Furthermore, the hybrid sensing approach incorporating QA and QD can be extended 
to generalized Ramsey interferometry
with one-axis twisting interactions~\cite{zollerPRX2021}, 
making it broadly applicable to diverse quantum systems such as Bose-Einstein condensates~\cite{mao2023quantum}, trapped ions~\cite{ion_science}, and Rydberg atom arrays~\cite{browaeys2020many,wu2021concise}.
Overall, our method offers a promising advancement in entanglement-enhanced metrology, with the potential to significantly improve applications such as the long-term stability of atomic clocks and other precision measurement systems~\cite{Ye24essay}.

{\it Acknowledgement.} 
We thank Leon Zaporski, Gustavo Velez, Guo-Qing Wang and Zhi-Yao Hu for helpful discussion. The following funding sources were acknowledged solely by the respective co-authors and are not related to any institutional support from MIT: M.X. is supported by National Natural Science Foundation of China (No.~12304543) and the Innovation Program for Quantum Science and Technology (No.~2021ZD0302100). L.-N.W. acknowledges support by the Hainan Province Science and Technology Talent Innovation Project (No.~KJRC2023L05). D.V.V. acknowledges funding by the Austrian Science Fund (FWF) [10.55776/COE1]. The primary numerical calculations in this study was carried out by Q.L. at MIT. The collaboration with non-MIT authors was limited to scholarly discussions and exchange of academic ideas. There was no transfer of materials, financial support, or institutional resource sharing between MIT and the non-US institutes involved in the collaboration.

{\it Data availability}---The data that support the findings of
this Letter are openly available~\cite{dataset}.

Q.L. and M.X. contributed equally to this work.

\bibliography{ref}

\onecolumngrid 
\clearpage
\includepdf[pages={{},-},
  ]{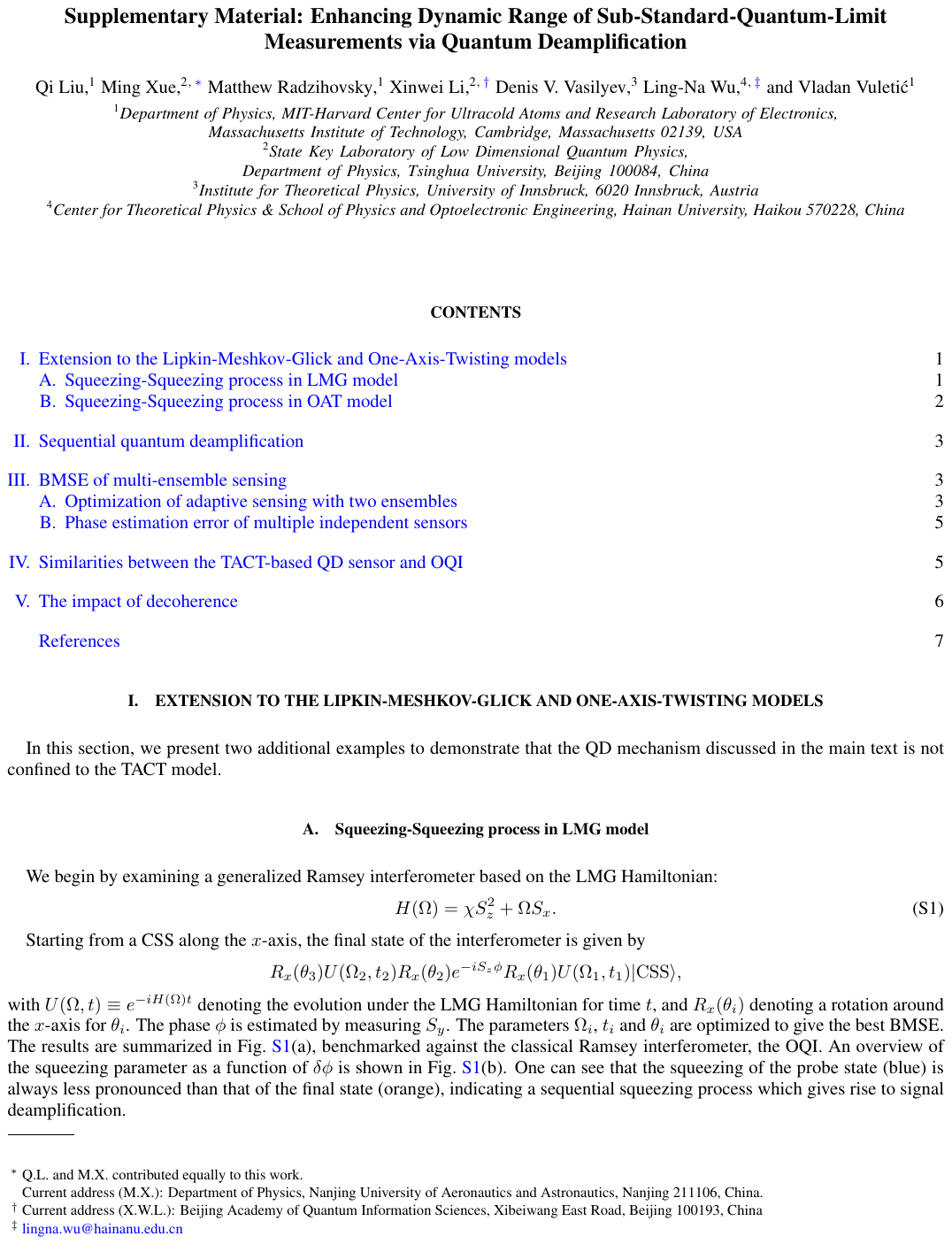}
\let\clearpage\relax
\end{document}